\documentclass[aps,prl,twocolumn,showpacs,superscriptaddress,groupedaddress]{revtex4}  
\usepackage{dcolumn}   
\usepackage{bm}        
\usepackage{amssymb}   
\usepackage{amsfonts}  
\usepackage{amsmath}   
\usepackage{mathcomp}  
\usepackage{graphicx}  
\usepackage{subfigure} 
\usepackage{multirow}  
\usepackage{hyperref}  

\usepackage{verbatim}
\usepackage{color}

\begin{document}

\title{Path Integral Monte Carlo Simulation of the Warm-Dense Homogeneous Electron Gas}

\author{Ethan W. Brown}
\email{brown122@illinois.edu}
\affiliation{Department of Physics, University of Illinois at Urbana-Champaign, 1110 W.\ Green St.\ , Urbana, IL  61801-3080, USA}
\affiliation{Lawrence Livermore National Lab, 7000 East Ave, L-415, Livermore CA 94550, USA}
\author{Bryan K. Clark}
\affiliation{Princeton Center For Theoretical Science, Princeton University, Princeton, NJ 08544}
\affiliation{Station Q, Microsoft Research, Santa Barbara, CA 93106, USA}
\author{Jonathan L. DuBois}
\affiliation{Lawrence Livermore National Lab, 7000 East Ave, L-415, Livermore CA 94550, USA}
\author{David M. Ceperley}
\affiliation{Department of Physics, University of Illinois at Urbana-Champaign, 1110 W.\ Green St.\ , Urbana, IL  61801-3080, USA}
\date{\today}

\begin{abstract}
We perform calculations of the $3D$ finite-temperature homogeneous electron gas (HEG) in the warm-dense regime ($r_{s} \equiv (3/4\pi n)^{1/3} a_{B}^{-1} = 1.0-40.0$ and $\Theta \equiv T/T_{F} = 0.0625-8.0$) using restricted path integral Monte Carlo (RPIMC). Precise energies, pair correlation functions, and structure factors are obtained. For all densities, we find a significant discrepancy between the ground state parameterized local density approximation (LDA) and our results around $T_{F}$. These results can be used as a benchmark for improved functionals, as well as input for orbital-free DFT formulations.
\end{abstract}

\maketitle

The one-component plasma (OCP),  a fundamental many body model, consists of a single species of charged particles immersed in a rigid neutralizing background. For electrons, the OCP is a model of simple metals and is often called the homogeneous electron gas (HEG), electron gas, or \emph{jellium}. At zero-temperature, it is customary to define the natural length scale $r_{s} a_{B} \equiv (3/4\pi n)^{1/3}$ and energy scale $Ry = e^{2}/2 a_{B}$. When $r_{s}$, the \emph{Wigner-Seitz radius}, is small (high density) ($r_{s} \rightarrow 0$), the kinetic energy term dominates and the system becomes qualitatively similar to a non-interacting gas. At low density ($r_{s} \rightarrow \infty$), the potential energy dominates and the system is predicted to form a Wigner crystal \cite{PhysRev.46.1002}. In 3D at intermediate densities, a partially polarized state is predicted to emerge \cite{BLOCH,PhysRevE.66.036703}.

Over the past few decades very accurate zero-temperature quantum Monte Carlo (QMC) calculations of the ground state HEG examined each of these phases \cite{PhysRevLett.45.566,PhysRevB.18.3126}. In addition to determining phase boundaries, the results of these studies have proven invaluable in the rigorous parameterization of functionals in ground state density functional theory (DFT) \cite{RevModPhys.80.3}. 

Recently there has been intense interest in extending the success of ground-state DFT to finite-temperature systems such as stellar, planetary interiors and other hot dense plasmas \cite{0953-8984-14-40-307,0741-3335-47-12B-S31,CTPP:CTPP239}. However, such attempts have met both fundamental and technical barriers when electrons have significant correlations.

Some of the first Monte Carlo simulations explored the phases of the classical OCP \cite{PhysRevA.8.3110}; note that its equation of state depends only on a single variable, the \emph{Coulomb coupling parameter} $\Gamma \equiv q^{2}/(r_{s}k_{B}T)$. First-order quantum mechanical effects have since been included \cite{1978PhyA...91..152J,1975PhLA...53..187H}. However, the accuracy of these results quickly deteriorate as the temperature is lowered and quantum correlations play a greater role \cite{PhysRevLett.76.4572}. This breakdown is most apparent in the \emph{warm-dense} regime where both $\Gamma$ and the \emph{electron degeneracy parameter} $\Theta \equiv T/T_{F}$ are close to unity.

Finite-temperature formulations of DFT have also met with challenges. There are two braod approaches to building finite-temperature functionals. In one approach, temperature effects are introduced by smearing the electronic density of states over a Fermi-Dirac distribution. As temperature increases, an ever-increasing number of molecular (Kohn-Sham) orbitals is required in order to evaluate the functional, making DFT calculations computationally intractable. In addition, although a useful approximation, this approach is not exact even in the limit of the exact ground state exchange functional as the Kohn-Sham orbitals need have no relation to the true excited states. A second approach is to use Orbital-Free Density Functional Theory (OFDFT) where the usual Kohn-Sham orbitals are replaced by another functional for the kinetic energy term \cite{2011arXiv1106.4792S,PhysRevB.84.075146}. However, an a priori way to determine such a functional has yet to materialize. Without a reliable benchmark, OFDFT is left to rely on Thomas-Fermi-like approximations which can incur errors an order of magnitude larger than typical DFT errors \cite{Karasiev20122519}. Having accurate finite-temperature energies for the $3D$ $HEG$ will help parametrize finite-temperature functionals.

In this paper, we provide accurate, first-principles thermodynamic data of the $3D$ HEG throughout the warm-dense regime, making firm connections to both previous semi-classical and ground-state studies. We utilize the Restricted Path Integral Monte Carlo (RPIMC) method. For a complete review of bosonic PIMC and its extension to fermions we refer the reader to \citep{RevModPhys.67.279} and \citep{binder1996monte,springerlink:10.1007/BF02181259}, respectively. Here we will only touch on parts of the method which are significant to this study.

PIMC allows in-principle exact calculations of equilibrium properties of quantum systems. For fermions, however, statistical weights of approximately equal magnitude and opposite sign make direct simulation computationally intractable at low temperatures. To circumvent this difficulty, a constraint is imposed such that sampled paths remain within the stricitly positive region of a trial density matrix; here we employ the free-electron density matrix,
\begin{equation}
  \rho_{0}(R,R',\tau) = (4\pi \tau/r_{s}^{2})^{-dN/2} \exp[-\frac{(r_{i}-r_{j}')^{2}}{4 \tau/r_{s}^{2}}]
  \label{eq:rhofree}
\end{equation}
We expect this approximation to be best at high temperature and at low-density when correlation effects are weak. Specifically we compare Eq. \ref{eq:rhofree} to the Feynman-Kac formulation for the full density matrix,
\begin{equation}
  \rho_{F}(R,R',\beta) = \mathcal{A} \rho_{0}(R,R',\beta) \langle \exp[-\int_{0}^{\beta}d\tau V(R(\tau))] \rangle
  \label{eq:rhoFK}
\end{equation}
where $\mathcal{A}$ is the anti-symmetrization operator and $\langle \dots \rangle$ denotes an average over Brownian walks from $R'$ to $R$. As $\beta \rightarrow 0$, this average tends to unity, leaving only the anti-symmetrized kinetic term. Thus for any potential $V(R)$ bounded from below, the nodes of the full density matrix equal the nodes of the free-particle density matrix in the high-temperature limit.

Furthermore, we expect free-particle nodes to be accurate for a homogeneous system, such as the electron gas, where translational symmetry constrains the possible nodal surfaces \cite{PhysRevLett.69.331}. Nevertheless further accounting of this approximation will be made through connection to prior semi-classical and ground-state simulations as well as exact evaluation of the unrestricted density matrix at higher temperatures.

We utilize the \emph{pair product approximation} to write the many-body density matrix as a product of high-temperature two-body density matrices. To account for the long-range nature of the Coulomb interaction, we split the density matrix into a short-range and long-range piece. Each short-range two-body density matrix is exactly solved at an even higher temperature, and then squared down to the temperature of interest $\tau^{-1}$. The long-range piece is then included via Ewald summation.

Rebuilding the many-body density matrix out of such two-body density matrices comes with an error that scales as $\sim \tau^{3}/r_{s}^{2}$. A more dominate form of time step error originates from paths which cross the nodal constraint in a time less than $\tau$. To help alleviate this effect, we use an image action to discourage paths from getting too close to nodes. An example of $\tau$ convergence is given in Supplementary Material along with the time steps used at all densities for both the fully spin-polarized ($\xi=1$) and unpolarized ($\xi=0$) systems.

For the fully spin-polarized system, we simulated $N=33$ particles, while for the unpolarized system, we simulated $N=66$ particles. Both $N$'s are so-called \emph{magic numbers} which completely fill a fixed number of bands for the free Fermi gas, helping to alleviate shell effects arising from a sharp Fermi surface. To further account for the finite-size of the simulation cell, we use the exact analytic correction for the ground-state homogeneous electron gas \cite{PhysRevLett.97.076404}. At intermediate and high densities, a second order correction to the kinetic energy is necessary \cite{PhysRevB.78.125106}, giving,
\begin{eqnarray*}
  \Delta T_{N} &=& \frac{1}{N}(\frac{\omega_{p}}{2} + \frac{5.264}{\pi r_{s}^{2} (2N)^{1/3}}[(1+\xi)^{2/3} + (1-\xi)^{2/3}]) \\
  \Delta V_{N} &=& \frac{\omega_{p}}{2N} \text{, } \Delta E_{N} = \Delta V_{N} + \Delta T_{N}
\end{eqnarray*}
where $\omega_{p} \equiv \sqrt{\frac{3}{r_{s}^{3}}}$ is the RPA plasmon frequency. At finite-temperature this correction is multiplied by $\tanh(\beta \omega_{p})$. Since it relies on the validity of the random phase approximation (RPA) at long-wavelength, this correction should still be accurate provided the small $k$ behavior of the static structure factor behaves as in the RPA. In Fig. \ref{fig:SF}, we verify this feature for the unpolarized state at $r_{s}=1.0$ and $r_{s}=10.0$. Note that for $\xi=0$, $S(k) = S_{\uparrow \uparrow}(k) + S_{\uparrow \downarrow}(k)$. Computed finite-size corrections are given in the Supplementary Material.

An additional error comes from the sampling error of the Monte Carlo algorithm itself. This error can be controlled by simply gathering more statistics through sampling additional configurations. Typical simulations required $\sim 10^{6}$ independent configurations for the statistical error to be on the same order as the other errors.

\begin{figure}
  \centering
  \includegraphics[width=\columnwidth]{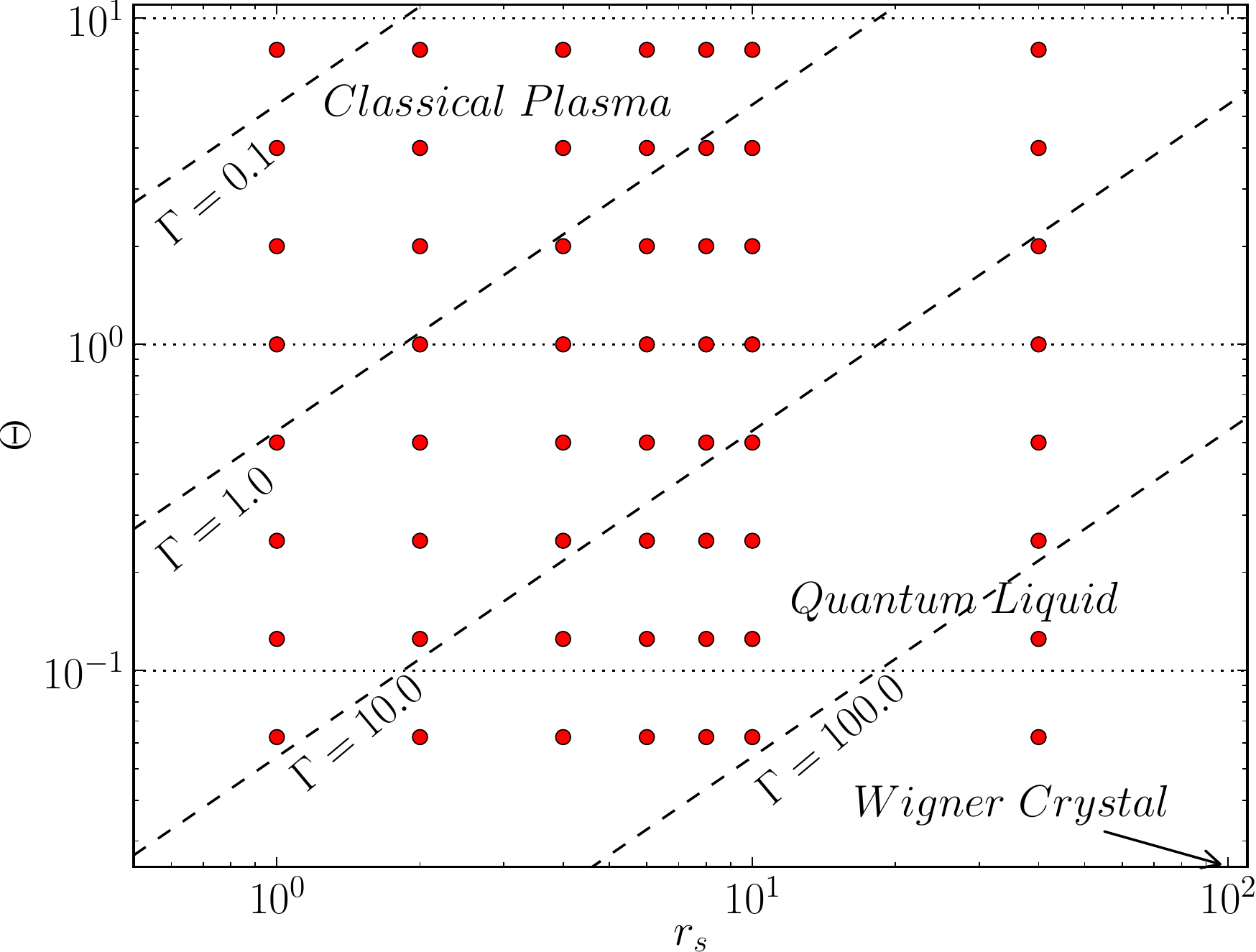}
  \caption{(color online) Temperature-Density points considered in the current study (dots). Several values of the Coulomb coupling parameter $\Gamma$ (dashed lines) and the electron degeneracy parameter $\Theta$ (dotted lines) are also shown.}
  \label{fig:PhaseDiagram}
\end{figure}


\begin{figure}[t]
  \centering
  \includegraphics[width=\columnwidth]{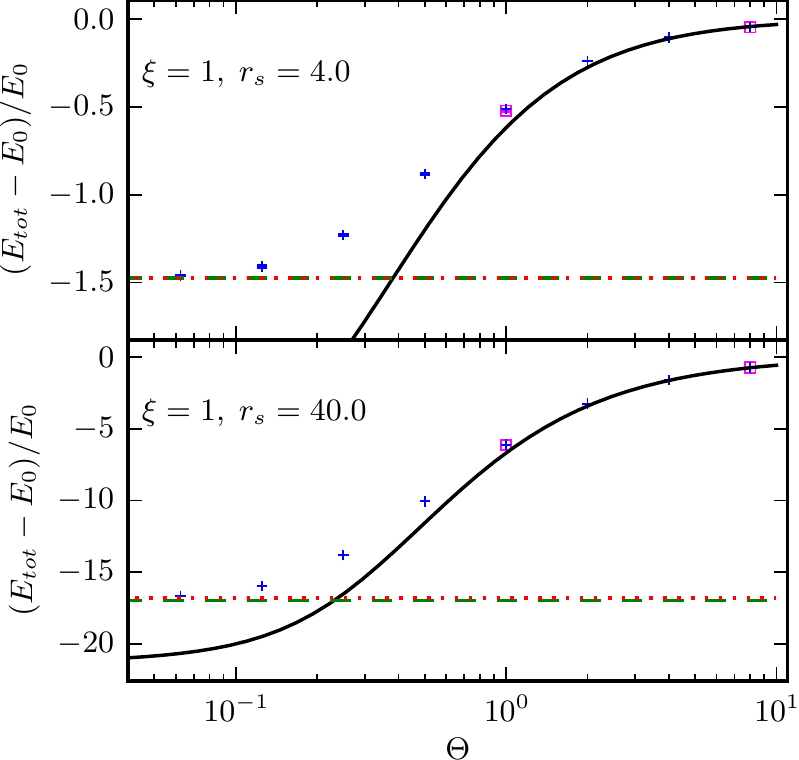}
  \caption{(color online) Excess energies for $r_{s}=4.0$ (top) and $r_{s}=40.0$ (bottom) for the polarized state. For both densities, the high temperature results fall smoothly on top of previous Monte Carlo energies for the classical electron gas \cite{PhysRevA.8.3110} (solid line). Differences from the classical coulomb gas occur for $\Theta < 2.0$ for $r_{s}=4.0$ and $\Theta<4.0$ for $r_{s}=40.0$. Simulations with the Fermion sign (squares) confirm the fixed-node results at $\Theta=1.0$ and $8.0$. The zero-temperature limit (dotted line) smoothly extrapolates to the ground-state QMC results of Ceperley-Alder \cite{PhysRevLett.45.566} (dashed line).}
  \label{fig:EvToTF}
\end{figure}

\begin{figure}[t]
  \centering
  \includegraphics[width=\columnwidth]{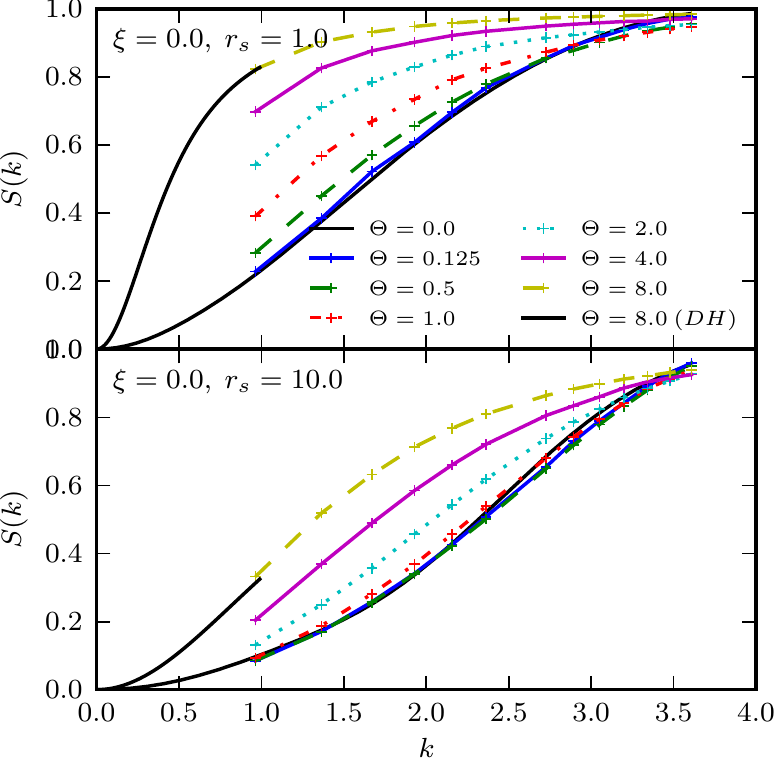}
  \caption{(color online) Static structure factors for $r_{s}=1.0$ and $r_{s}=10.0$ in the unpolarized state. At $\Theta=0.0$ we plot the ground state structure factor from Ref. \cite{PhysRevB.61.7353}. Also shown is the small $k$ part of $S_{DH}(k)$ at $\Theta=8.0$, see Eq. \ref{eq:DHSF}. 
}
  \label{fig:SF}
\end{figure}

\begin{figure}[t]
  \centering
  \includegraphics[width=\columnwidth]{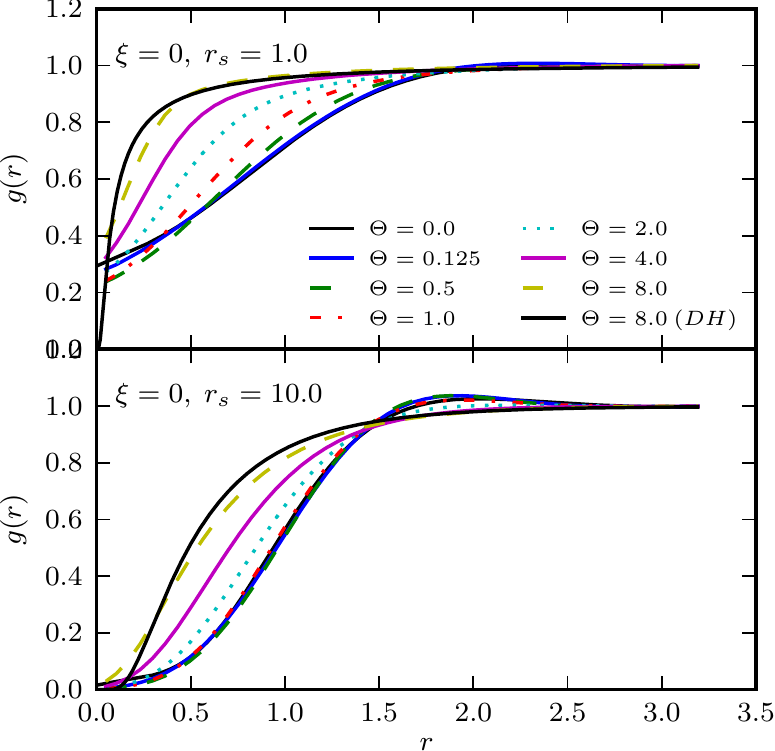}
  \caption{(color online) Pair correlation functions for $r_{s}=1.0$ and $r_{s}=10.0$ in the unpolarized state. At $\Theta=0.0$ is shown the ground state correlation function from Ref. \cite{PhysRevB.61.7353}. Deviation from RPIMC is seen at small $r$, but this is most likely due to poor ground-state QMC data \cite{PhysRevB.66.165118}. Also shown is the small $r$ part of $g_{DH}(r)$ at $\Theta=8.0$, see Eq. \ref{eq:DHPC}. The Debye-Huckel limit is not yet reached at $\Theta=8.0$ for the lower density $r_{s}=10.0$.}
  \label{fig:PC}
\end{figure}

We have calculated  energies, pair correlation functions, and structure factors of the $3D$ HEG for densities ranging from $r_{s}=1.0$ to $40.0$ and temperatures ranging from $\Theta =0.0625$ to $8.0$ as shown in Figure (1). At each density, we observe a smooth convergence to previous semi-classical studies \cite{PhysRevA.8.3110} at high temperature.

In Fig. \ref{fig:EvToTF} we plot the total excess energy for the polarized system at all temperatures with $r_{s}=4.0$ and $40.0$. At the highest temperatures, our results match well with the purely classical Monte Carlo results of Ref. \cite{PhysRevA.8.3110} (solid line). For a few select points, we have performed the much more time-consuming but more accurate, signful PIMC simulation (squares). These points which are essentially exact, i.e. without possible nodal error, match well with fixed-node results, see Supplementary Material. Finally, we know from Fermi liquid theory the low-temperature gas should have a linear form for the heat capacity, and therefore a quadratic form for the internal energy. Thus for each density we fit the low-temperature points to a quadratic function and extrapolate to $0K$. Fig. \ref{fig:EvToTF} shows the extrapolated results (dotted line) match well with the zero-temperature QMC results of Ceperley-Alder \cite{PhysRevLett.45.566} (dashed line). For precise values see the Supplementary Material.

Fig. \ref{fig:SF} shows the calculated structure factors for the unpolarized state at $r_{s}=1.0$ and $r_{s}=10.0$. At all densities and polarizations, we see a smooth convergence to both the ground-state and classical Debye-Huckel limits. Zero-temperature curves are generated through an analytic fit to previous QMC data \cite{PhysRevB.61.7353}, while Debye-Huckel curves are generated using \cite{PhysRevA.8.3096},
\begin{equation}
  S_{DH}(k) = \frac{k^{2}}{k^{2} + 3 \Gamma}
  \label{eq:DHSF}
\end{equation}

Fig. \ref{fig:PC} shows the pair correlation functions for the same systems. Again, we see a convergence to analytic ground-state curves. The small $r$ behavior slightly deviates for $r_{s}=1.0$, but this is due to the poor quality of small $r$ QMC data which was used to create the analytic fit \cite{PhysRevA.8.3096}. We also plot the Debye-Huckel pair correlation function given by,
\begin{equation}
  g_{DH}(r) = exp[-(r\theta) exp(-r (3/\theta)^{\frac{1}{2}})]
  \label{eq:DHPC}
\end{equation}
where $\theta \equiv \frac{r_{s}T}{2}$. As was noted in Ref. \cite{PhysRevA.8.3096}, convergence of $g(r)$ to the Debye-Huckel limit is slower than for the corresponding $S(k)$.

Through this comparison of our results against existing numerical and analytical data, we conclude the free-particle nodal approximation performs well for the densities studied. Further investigation is needed at even smaller values of $r_{s}$ and lower temperatures in order to determine precisely where this approximation begins to fail. Such studies will necessarily require algorithmic improvements, however, because of difficulty in sampling paths at high density and low temperature \cite{PhysRevLett.69.331}.

Finally, we have evaluated the exchange-correlation energy $e_{xc}$, an essential quantity in any DFT formulation, defined
\begin{equation}
  E_{xc}(T) \equiv E_{tot}(T) - E_{0}(T)
\end{equation}
where $E_{0}$ is the kinetic energy of a free Fermi gas at temperature $T$. As is customary, we further break up $E_{xc}$ into exchange and correlation parts,
\begin{equation}
  E_{xc}(T) = E_{x}(T) + E_{c}(T)
\end{equation}
where $E_{x}(T)$ is the Hartree-Fock exchange energy for a free Fermi gas at temperature $T$.

\begin{figure}[t]
  \centering
  \includegraphics[width=\columnwidth]{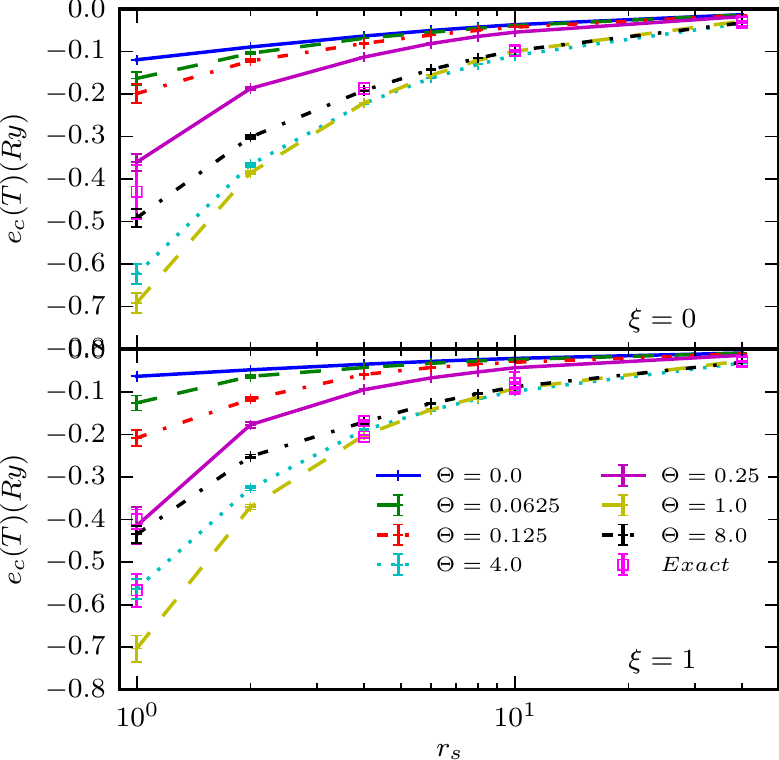}
  \caption{(color online) Correlation energy $e_{c}(T)$ of the $3D$ HEG at several temperatures and densities for the unpolarized (top) and fully spin-polarized (bottom) states. \emph{Exact} (signful) calculations (squares) confirm the fixed-node results where possible. Only shown are those points which had reasonable error bars. See the Supplementary Material for the rest of the data points.}
  \label{fig:Ec}
\end{figure}

By calculating $E_{tot}(T)$ through RPIMC simulations we were able to determine $E_{c}(T)$ at all studied densities for both the fully spin-polarized and unpolarized states. As one can see in Fig. \ref{fig:Ec}, correlation effects increase both with density (smaller $r_{s}$) and temperature up to a temperature above the Fermi temperature $T_{F}$. Above this temperature, the electron gas begins to be less correlated. This represents the point at which electron screening is a dominant effect, the interaction becomes effectively short-ranged, and the Debye-approximation becomes relatively accurate \cite{PhysRevA.8.3096}. As the density increases, the value of $\Theta$ at which this occurs decreases. At $r_{s}=1.0$ the maximal effect of interactions occurs very near $T_{F}$, $\Theta=1$.

In conclusion we have used RPIMC with free-particle nodes to calculate energies, pair correlation factors, and structure factors for the $3D$ HEG throughout the warm-dense regime. Systematic errors, including finite-size effects, time-step, and statistical fluctuations, are controlled for. Through cross-validation with previous ground-state and classical QMC and exact finite temperature calculations, we estimate that bias from the use of the free particle density matrix in the constraint is small for the density/temperature points simulated. This does not exclude the possibility of fixed-node error at higher densities and lower temperatures. In future work we will quantify this error by finding better nodal structures and doing calculations without such uncontrolled approximations.

For those who wish to use this data as a benchmark, we are providing all data both in the Supplementary Material at [URL will be inserted by publisher] and a repository hosted at \url{http://github.com/3dheg/3DHEG}. Instructions on how to access and use the data are available at that address.

We would like to thank Jeremy McMinis, Norm Tubman, David ChangMo Yang, Miguel Morales, and Markus Holzmann for useful discussions. This work was supported by grant DE-FG52-09NA29456.  In addition, the work of E. Brown and J. DuBois was performed under the auspices of the U.S. Department of Energy by Lawrence Livermore National Laboratory under Contract DE-AC52-07NA27344 with support from LDRD 10-ERD-058 and the Lawrence Scholar program.  Computational resources included Jaguar and Kraken at Oak Ridge National Laboratory through XSEDE, and LC machines at Lawrence Livermore National Laboratory through the institutional computation grand challenge program.

\bibliography{prl}{}

\pagebreak

\begin{figure}
  \centering
  \includegraphics[width=\columnwidth]{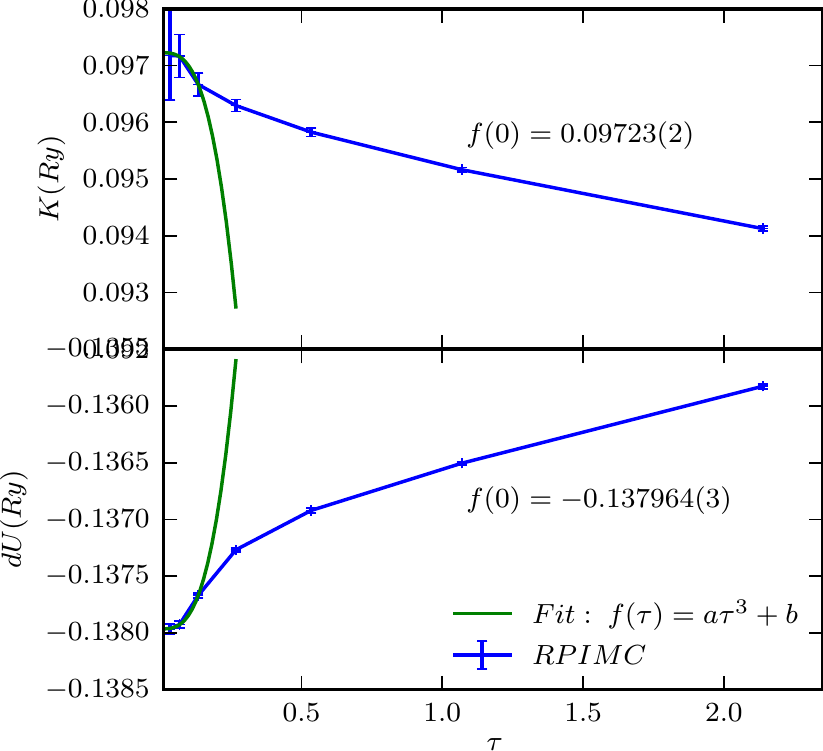}
  \caption{(color online) Convergence in $\tau$ for $\xi = 1$ and $r_{s} = 10.0$.}
  \label{fig:timestep}
\end{figure}

\begin{table}
  \begin{tabular}{cc|c|c|c|c|c|c|l}
    \cline{2-8}
    \multicolumn{1}{c|}{} & \multicolumn{7}{c|}{$r_{s}$} \\ \cline{2-8}
    \multicolumn{1}{c|}{} & $1.0$ & $2.0$ & $4.0$ & $6.0$ & $8.0$ & $10.0$ & $40.0$ \\ \cline{1-8}
    \multicolumn{1}{|c|}{$\xi=0$} & $0.0005$ & $0.0042$ & $0.0170$ & $0.0382$ & $0.0679$ & $0.1061$ & $3.3938$ \\ \cline{1-8}
    \multicolumn{1}{|c|}{$\xi=1$} & $0.0007$ & $0.0027$ & $0.0214$ & $0.0481$ & $0.0855$ & $0.1336$ & $2.1380$ \\ \cline{1-8}
  \end{tabular}
  \caption{Time step $\tau$ ($Ry^{-1}$) used for each density $r_{s}$ and polarization $\xi$.}
  \label{tab:timesteps}
\end{table}

\begin{table}
  \begin{tabular}{|c|c|c|c|c|c|c|c|c|c|}
    \cline{1-10}
    $r_{s}$ & $\Theta$ & $K$ & $V$ & $\Delta K_{N}$ & $\Delta V_{N}$ & $E_{tot}$ & $E_{0}$ & $E_{x,HF}$ & $E_{c}$ \\
    \cline{1-10}
   $1.0$ & $0.0625$ & $2.27(1)$ & $-1.0839(3)$ & $0.021257$ & $0.026226$ & $1.18(1)$ & $2.245191$ & $-0.896908$ & $-0.16(1)$ \\
    \cline{1-10}
   $2.0$ & $0.0625$ & $0.598(2)$ & $-0.5892(1)$ & $0.008032$ & $0.009277$ & $0.008(2)$ & $0.561298$ & $-0.448454$ & $-0.104(2)$ \\
    \cline{1-10}
   $4.0$ & $0.0625$ & $0.1678(6)$ & $-0.32062(5)$ & $0.002969$ & $0.003280$ & $-0.1528(6)$ & $0.140324$ & $-0.224227$ & $-0.0689(6)$ \\
    \cline{1-10}
   $6.0$ & $0.0625$ & $0.0816(2)$ & $-0.22389(2)$ & $0.001647$ & $0.001786$ & $-0.1423(2)$ & $0.062366$ & $-0.149485$ & $-0.0551(2)$ \\
    \cline{1-10}
   $8.0$ & $0.0625$ & $0.0506(1)$ & $-0.17313(1)$ & $0.001082$ & $0.001160$ & $-0.1225(1)$ & $0.035081$ & $-0.112113$ & $-0.0455(1)$ \\
    \cline{1-10}
   $10.0$ & $0.0625$ & $0.03524(8)$ & $-0.141605(9)$ & $0.000780$ & $0.000830$ & $-0.10637(9)$ & $0.022452$ & $-0.089691$ & $-0.03913(9)$ \\
    \cline{1-10}
   $40.0$ & $0.0625$ & $0.004221(2)$ & $-0.039362(1)$ & $0.000101$ & $0.000104$ & $-0.035141(3)$ & $0.001403$ & $-0.022423$ & $-0.014121(3)$ \\
    \cline{1-10}
   $1.0$ & $0.125$ & $2.37(2)$ & $-1.0783(6)$ & $0.021257$ & $0.026209$ & $1.29(2)$ & $2.348227$ & $-0.855560$ & $-0.20(2)$ \\
    \cline{1-10}
   $2.0$ & $0.125$ & $0.626(3)$ & $-0.5890(2)$ & $0.008032$ & $0.009276$ & $0.037(3)$ & $0.587057$ & $-0.427780$ & $-0.122(3)$ \\
    \cline{1-10}
   $4.0$ & $0.125$ & $0.1727(6)$ & $-0.32088(5)$ & $0.002969$ & $0.003280$ & $-0.1482(7)$ & $0.146764$ & $-0.213890$ & $-0.0811(7)$ \\
    \cline{1-10}
   $6.0$ & $0.125$ & $0.0854(2)$ & $-0.22412(2)$ & $0.001647$ & $0.001786$ & $-0.1387(3)$ & $0.065229$ & $-0.142593$ & $-0.0613(3)$ \\
    \cline{1-10}
   $8.0$ & $0.125$ & $0.0527(1)$ & $-0.17339(1)$ & $0.001082$ & $0.001160$ & $-0.1207(1)$ & $0.036691$ & $-0.106945$ & $-0.0504(1)$ \\
    \cline{1-10}
   $10.0$ & $0.125$ & $0.0368(1)$ & $-0.14184(1)$ & $0.000780$ & $0.000830$ & $-0.1050(1)$ & $0.023482$ & $-0.085556$ & $-0.0430(1)$ \\
    \cline{1-10}
   $40.0$ & $0.125$ & $0.004336(5)$ & $-0.039408(1)$ & $0.000101$ & $0.000104$ & $-0.035072(5)$ & $0.001468$ & $-0.021389$ & $-0.015151(5)$ \\
    \cline{1-10}
   $1.0$ & $0.25$ & $2.68(1)$ & $-1.0738(6)$ & $0.021257$ & $0.026175$ & $1.60(1)$ & $2.713435$ & $-0.748509$ & $-0.36(1)$ \\
    \cline{1-10}
   $2.0$ & $0.25$ & $0.708(2)$ & $-0.5909(2)$ & $0.008032$ & $0.009274$ & $0.117(3)$ & $0.678359$ & $-0.374254$ & $-0.187(3)$ \\
    \cline{1-10}
   $4.0$ & $0.25$ & $0.1913(7)$ & $-0.32206(6)$ & $0.002969$ & $0.003280$ & $-0.1307(7)$ & $0.169590$ & $-0.187127$ & $-0.1132(7)$ \\
    \cline{1-10}
   $6.0$ & $0.25$ & $0.0943(4)$ & $-0.22517(3)$ & $0.001647$ & $0.001786$ & $-0.1309(4)$ & $0.075373$ & $-0.124751$ & $-0.0815(4)$ \\
    \cline{1-10}
   $8.0$ & $0.25$ & $0.0583(2)$ & $-0.17416(2)$ & $0.001082$ & $0.001160$ & $-0.1158(2)$ & $0.042397$ & $-0.093564$ & $-0.0647(2)$ \\
    \cline{1-10}
   $10.0$ & $0.25$ & $0.0399(1)$ & $-0.14234(1)$ & $0.000780$ & $0.000830$ & $-0.1024(1)$ & $0.027134$ & $-0.074851$ & $-0.0547(1)$ \\
    \cline{1-10}
   $40.0$ & $0.25$ & $0.00456(1)$ & $-0.039471(2)$ & $0.000101$ & $0.000104$ & $-0.03491(1)$ & $0.001696$ & $-0.018713$ & $-0.01789(1)$ \\
    \cline{1-10}
   $1.0$ & $0.5$ & $3.69(2)$ & $-1.0494(8)$ & $0.021234$ & $0.026135$ & $2.64(2)$ & $3.762881$ & $-0.557716$ & $-0.57(2)$ \\
    \cline{1-10}
   $2.0$ & $0.5$ & $0.948(4)$ & $-0.5835(2)$ & $0.008031$ & $0.009270$ & $0.365(4)$ & $0.940720$ & $-0.278858$ & $-0.297(4)$ \\
    \cline{1-10}
   $4.0$ & $0.5$ & $0.2494(9)$ & $-0.32189(6)$ & $0.002969$ & $0.003280$ & $-0.0725(9)$ & $0.235180$ & $-0.139429$ & $-0.1683(9)$ \\
    \cline{1-10}
   $6.0$ & $0.5$ & $0.1182(4)$ & $-0.22525(3)$ & $0.001647$ & $0.001786$ & $-0.1071(4)$ & $0.104524$ & $-0.092953$ & $-0.1186(4)$ \\
    \cline{1-10}
   $8.0$ & $0.5$ & $0.0705(2)$ & $-0.17432(2)$ & $0.001082$ & $0.001160$ & $-0.1039(2)$ & $0.058795$ & $-0.069714$ & $-0.0929(2)$ \\
    \cline{1-10}
   $10.0$ & $0.5$ & $0.0477(1)$ & $-0.14252(1)$ & $0.000780$ & $0.000830$ & $-0.0948(1)$ & $0.037629$ & $-0.055772$ & $-0.0766(1)$ \\
    \cline{1-10}
   $40.0$ & $0.5$ & $0.004879(6)$ & $-0.039473(1)$ & $0.000101$ & $0.000104$ & $-0.034594(7)$ & $0.002352$ & $-0.013943$ & $-0.023003(7)$ \\
    \cline{1-10}
   $1.0$ & $1.0$ & $6.16(2)$ & $-0.9468(9)$ & $0.020292$ & $0.027206$ & $5.21(2)$ & $6.249377$ & $-0.347701$ & $-0.69(2)$ \\
    \cline{1-10}
   $2.0$ & $1.0$ & $1.547(3)$ & $-0.5442(3)$ & $0.007954$ & $0.009352$ & $1.003(4)$ & $1.562344$ & $-0.173851$ & $-0.385(4)$ \\
    \cline{1-10}
   $4.0$ & $1.0$ & $0.3914(9)$ & $-0.3086(1)$ & $0.002966$ & $0.003283$ & $0.083(1)$ & $0.390586$ & $-0.086925$ & $-0.221(1)$ \\
    \cline{1-10}
   $6.0$ & $1.0$ & $0.1784(4)$ & $-0.21883(5)$ & $0.001647$ & $0.001786$ & $-0.0404(4)$ & $0.173594$ & $-0.057950$ & $-0.1560(4)$ \\
    \cline{1-10}
   $8.0$ & $1.0$ & $0.1037(2)$ & $-0.17051(3)$ & $0.001082$ & $0.001160$ & $-0.0668(2)$ & $0.097647$ & $-0.043463$ & $-0.1210(2)$ \\
    \cline{1-10}
   $10.0$ & $1.0$ & $0.0681(1)$ & $-0.14003(2)$ & $0.000780$ & $0.000830$ & $-0.0719(1)$ & $0.062494$ & $-0.034770$ & $-0.0996(1)$ \\
    \cline{1-10}
   $40.0$ & $1.0$ & $0.005817(5)$ & $-0.039321(1)$ & $0.000101$ & $0.000104$ & $-0.033504(7)$ & $0.003906$ & $-0.008693$ & $-0.028717(7)$ \\
    \cline{1-10}
   $1.0$ & $2.0$ & $11.49(2)$ & $-0.800(1)$ & $0.015634$ & $0.034942$ & $10.69(2)$ & $11.566451$ & $-0.191763$ & $-0.69(2)$ \\
    \cline{1-10}
   $2.0$ & $2.0$ & $2.871(4)$ & $-0.4750(4)$ & $0.006982$ & $0.010635$ & $2.396(4)$ & $2.891613$ & $-0.095881$ & $-0.400(4)$ \\
    \cline{1-10}
   $4.0$ & $2.0$ & $0.717(1)$ & $-0.2795(1)$ & $0.002834$ & $0.003434$ & $0.438(1)$ & $0.722903$ & $-0.047941$ & $-0.237(1)$ \\
    \cline{1-10}
   $6.0$ & $2.0$ & $0.3217(5)$ & $-0.2025(1)$ & $0.001615$ & $0.001821$ & $0.1192(6)$ & $0.321290$ & $-0.031960$ & $-0.1701(6)$ \\
    \cline{1-10}
   $8.0$ & $2.0$ & $0.1824(2)$ & $-0.15972(6)$ & $0.001071$ & $0.001171$ & $0.0226(3)$ & $0.180726$ & $-0.023970$ & $-0.1341(3)$ \\
    \cline{1-10}
   $10.0$ & $2.0$ & $0.1179(1)$ & $-0.13253(3)$ & $0.000776$ & $0.000834$ & $-0.0146(1)$ & $0.115665$ & $-0.019176$ & $-0.1111(1)$ \\
    \cline{1-10}
   $40.0$ & $2.0$ & $0.008422(5)$ & $-0.038671(2)$ & $0.000101$ & $0.000104$ & $-0.030249(8)$ & $0.007229$ & $-0.004794$ & $-0.032684(8)$ \\
    \cline{1-10}
   $1.0$ & $4.0$ & $22.38(2)$ & $-0.6373(8)$ & $0.009319$ & $0.057374$ & $21.74(2)$ & $22.465711$ & $-0.099617$ & $-0.62(2)$ \\
    \cline{1-10}
   $2.0$ & $4.0$ & $5.593(4)$ & $-0.3923(3)$ & $0.004672$ & $0.015833$ & $5.200(4)$ & $5.616428$ & $-0.049808$ & $-0.366(4)$ \\
    \cline{1-10}
   $4.0$ & $4.0$ & $1.397(1)$ & $-0.2393(1)$ & $0.002183$ & $0.004455$ & $1.157(1)$ & $1.404107$ & $-0.024904$ & $-0.222(1)$ \\
    \cline{1-10}
   $6.0$ & $4.0$ & $0.6220(4)$ & $-0.1775(1)$ & $0.001348$ & $0.002181$ & $0.4445(5)$ & $0.624048$ & $-0.016603$ & $-0.1629(5)$ \\
    \cline{1-10}
   $8.0$ & $4.0$ & $0.3504(2)$ & $-0.14244(7)$ & $0.000940$ & $0.001334$ & $0.2080(3)$ & $0.351027$ & $-0.012452$ & $-0.1306(3)$ \\
    \cline{1-10}
   $10.0$ & $4.0$ & $0.2250(1)$ & $-0.11955(5)$ & $0.000704$ & $0.000919$ & $0.1054(2)$ & $0.224657$ & $-0.009962$ & $-0.1093(2)$ \\
    \cline{1-10}
   $40.0$ & $4.0$ & $0.014683(5)$ & $-0.037064(4)$ & $0.000100$ & $0.000104$ & $-0.022381(9)$ & $0.014041$ & $-0.002490$ & $-0.033932(9)$ \\
    \cline{1-10}
   $1.0$ & $8.0$ & $44.40(2)$ & $-0.4842(7)$ & $0.004908$ & $0.104221$ & $43.92(2)$ & $44.457612$ & $-0.050512$ & $-0.49(2)$ \\
    \cline{1-10}
   $2.0$ & $8.0$ & $11.098(4)$ & $-0.3107(2)$ & $0.002577$ & $0.028499$ & $10.788(4)$ & $11.114403$ & $-0.025256$ & $-0.302(4)$ \\
    \cline{1-10}
   $4.0$ & $8.0$ & $2.7702(9)$ & $-0.1956(1)$ & $0.001302$ & $0.007463$ & $2.575(1)$ & $2.778601$ & $-0.012628$ & $-0.191(1)$ \\
    \cline{1-10}
   $6.0$ & $8.0$ & $1.2320(4)$ & $-0.14807(7)$ & $0.000856$ & $0.003433$ & $1.0839(5)$ & $1.234934$ & $-0.008419$ & $-0.1426(5)$ \\
    \cline{1-10}
   $8.0$ & $8.0$ & $0.6933(2)$ & $-0.12087(5)$ & $0.000629$ & $0.001993$ & $0.5725(3)$ & $0.694650$ & $-0.006314$ & $-0.1159(3)$ \\
    \cline{1-10}
   $10.0$ & $8.0$ & $0.4442(1)$ & $-0.10277(4)$ & $0.000492$ & $0.001314$ & $0.3414(2)$ & $0.444576$ & $-0.005051$ & $-0.0981(2)$ \\
    \cline{1-10}
   $40.0$ & $8.0$ & $0.028112(9)$ & $-0.034306(8)$ & $0.000091$ & $0.000115$ & $-0.00619(1)$ & $0.027786$ & $-0.001263$ & $-0.03272(1)$ \\
    \cline{1-10}
  \end{tabular}
  \caption{Measured energies at all densities and temperatures simulated for the unpolarized ($\xi=0.0$) gas. Shown for each density $r_{s}$ and temperature $\Theta$ are the size-corrected values for the Kinetic $K$ and Potential $V$ energies, their respective finite-size corrections $\Delta K_{N}$ and $\Delta V_{N}$, the resulting total energy $E_{tot}$, the free electron energy $E_{0}$, the Hartree-Fock exchange energy $E_{x,HF}$, and the resulting correlation energy $E_{c}$.}
  \label{tab:energies-0}
\end{table}

\begin{table}
  \begin{tabular}{|c|c|c|c|c|}
    \cline{1-5}
    $r_{s}$ & $E_{tot}(0)$ & $\displaystyle \lim_{T\rightarrow 0} E_{tot}(T)$ & $E_{c}(0) $ & $\displaystyle \lim_{T\rightarrow 0} E_{c}(T)$ \\
    \cline{1-5}
    $1.0$ & $1.1726(2)^{b}$ & $1.16(1)$ & $-0.1210(2)$ & $-0.13(1)$ \\
    \cline{1-5}
    $2.0$ & $0.0041(4)^{a}$ & $0.003(2)$ & $-0.0902(4)$ & $-0.091(2)$ \\
    \cline{1-5}
    $4.0$ & $-0.1547(1)^{d}$ & $-0.1542(5)$ & $-0.0637(1)$ & $-0.0632(5)$ \\
    \cline{1-5}
    $6.0$ & $-0.1422(1)^{d}$ & $-0.1425(2)$ & $-0.0509(1)$ & $-0.0512(2)$ \\
    \cline{1-5}
    $8.0$ & $-0.1228(1)^{d}$ & $-0.1228(1)$ & $-0.0428(1)$ & $-0.0428(1)$ \\
    \cline{1-5}
    $10.0$ & $-0.10687(2)^{b}$ & $-0.10643(8)$ & $-0.03734(2)$ & $-0.03690(8)$ \\
    \cline{1-5}
    $40.0$ & $-0.0352375(6)^{c}$ & $-0.035153(3)$ & $-0.0137104(6)$ & $-0.013626(3)$ \\
    \cline{1-5}
  \end{tabular}
  \caption{Zero-temperature extrapolations, $\displaystyle \lim_{T\rightarrow 0} E_{tot}(T)$, of finite-temperature PIMC calculations for the unpolarized ($\xi=0.0$). We compare $E_{tot}(0)$ directly to previous QMC studies where possible (a, \cite{PhysRevLett.45.566}), (b, \cite{PhysRevE.66.036703}), (c, \cite{PhysRevB.58.6800}), otherwise the Perdew-Zunger parameterization (d, \cite{PhysRevB.23.5048}) is used.}
  \label{tab:E0-0}
\end{table}

\begin{table}
  \begin{tabular}{|c|c|c|c|c|}
    \cline{1-5}
    $r_{s}$ & $\Theta$ & $\langle sgn \rangle$ &  $E_{tot}^{exact}$ & $E_{tot}$ \\
    \cline{1-5}
    $1.0$ & $1.0$ & $0.0002(10)$ & $-17(123)$ & $5.21(2)$ \\
    \cline{1-5}
    $4.0$ & $1.0$ & $0.0040(7)$ & $0.08(16)$ & $0.083(1)$ \\
    \cline{1-5}
    $10.0$ & $1.0$ & $0.092(1)$ & $-0.071(2)$ & $-0.0719(1)$ \\
    \cline{1-5}
    $40.0$ & $1.0$ & $0.7689(9)$ & $-0.03350(3)$ & $-0.033504(7)$ \\
    \cline{1-5}
    $1.0$ & $8.0$ & $0.5568(5)$ & $43.98(6)$ & $43.92(2)$ \\
    \cline{1-5}
    $4.0$ & $8.0$ & $0.7415(3)$ & $2.579(2)$ & $2.575(1)$ \\
    \cline{1-5}
    $10.0$ & $8.0$ & $0.9070(2)$ & $0.3418(2)$ & $0.3414(2)$ \\
    \cline{1-5}
    $40.0$ & $8.0$ & $0.99826(3)$ & $-0.006202(8)$ & $-0.00619(1)$ \\
    \cline{1-5}
  \end{tabular}
  \caption{Comparison of signful calculation $E_{tot}^{exact}$ with the fixed-node calculations $E_{tot}$ for the unpolarized ($\xi=0.0$) gas at select densities and temperatures. The average value of the sign is shown for reference.}
  \label{tab:Exct-0}
\end{table}

\begin{table}
  \begin{tabular}{|c|c|c|c|c|c|c|c|c|c|}
    \cline{1-10}
    $r_{s}$ & $\Theta$ & $K$ & $V$ & $\Delta K_{N}$ & $\Delta V_{N}$ & $E_{tot}$ & $E_{0}$ & $E_{x,HF}$ & $E_{c}$ \\
    \cline{1-10}
   $1.0$ & $0.0625$ & $3.51(1)$ & $-1.2052(3)$ & $0.036656$ & $0.052315$ & $2.31(1)$ & $3.564018$ & $-1.130033$ & $-0.13(1)$ \\
    \cline{1-10}
   $2.0$ & $0.0625$ & $0.896(3)$ & $-0.6337(1)$ & $0.014599$ & $0.018546$ & $0.263(3)$ & $0.891004$ & $-0.565016$ & $-0.063(3)$ \\
    \cline{1-10}
   $4.0$ & $0.0625$ & $0.2304(6)$ & $-0.33271(6)$ & $0.005571$ & $0.006560$ & $-0.1023(7)$ & $0.222751$ & $-0.282508$ & $-0.0425(7)$ \\
    \cline{1-10}
   $6.0$ & $0.0625$ & $0.1077(2)$ & $-0.22960(3)$ & $0.003132$ & $0.003571$ & $-0.1219(3)$ & $0.099000$ & $-0.188339$ & $-0.0325(3)$ \\
    \cline{1-10}
   $8.0$ & $0.0625$ & $0.0638(2)$ & $-0.17631(2)$ & $0.002072$ & $0.002320$ & $-0.1125(2)$ & $0.055688$ & $-0.141254$ & $-0.0269(2)$ \\
    \cline{1-10}
   $10.0$ & $0.0625$ & $0.0426(1)$ & $-0.14358(1)$ & $0.001501$ & $0.001660$ & $-0.1010(1)$ & $0.035640$ & $-0.113003$ & $-0.0236(1)$ \\
    \cline{1-10}
   $40.0$ & $0.0625$ & $0.004561(8)$ & $-0.039431(2)$ & $0.000198$ & $0.000207$ & $-0.03487(1)$ & $0.002228$ & $-0.028251$ & $-0.00885(1)$ \\
    \cline{1-10}
   $1.0$ & $0.125$ & $3.64(1)$ & $-1.1961(5)$ & $0.036656$ & $0.052143$ & $2.44(1)$ & $3.727579$ & $-1.077938$ & $-0.21(1)$ \\
    \cline{1-10}
   $2.0$ & $0.125$ & $0.906(4)$ & $-0.6302(1)$ & $0.014599$ & $0.018535$ & $0.276(4)$ & $0.931895$ & $-0.538969$ & $-0.117(4)$ \\
    \cline{1-10}
   $4.0$ & $0.125$ & $0.237(1)$ & $-0.3318(1)$ & $0.005571$ & $0.006559$ & $-0.095(1)$ & $0.232974$ & $-0.269484$ & $-0.059(1)$ \\
    \cline{1-10}
   $6.0$ & $0.125$ & $0.1104(8)$ & $-0.22911(7)$ & $0.003132$ & $0.003571$ & $-0.1187(9)$ & $0.103544$ & $-0.179656$ & $-0.0426(9)$ \\
    \cline{1-10}
   $8.0$ & $0.125$ & $0.0648(5)$ & $-0.17617(4)$ & $0.002072$ & $0.002320$ & $-0.1113(5)$ & $0.058243$ & $-0.134742$ & $-0.0348(5)$ \\
    \cline{1-10}
   $10.0$ & $0.125$ & $0.0429(2)$ & $-0.14350(2)$ & $0.001501$ & $0.001660$ & $-0.1006(2)$ & $0.037276$ & $-0.107794$ & $-0.0301(2)$ \\
    \cline{1-10}
   $40.0$ & $0.125$ & $0.00455(1)$ & $-0.039429(2)$ & $0.000198$ & $0.000207$ & $-0.03488(1)$ & $0.002330$ & $-0.026948$ & $-0.01026(1)$ \\
    \cline{1-10}
   $1.0$ & $0.25$ & $4.12(4)$ & $-1.171(1)$ & $0.036651$ & $0.051808$ & $2.95(4)$ & $4.307310$ & $-0.943062$ & $-0.41(4)$ \\
    \cline{1-10}
   $2.0$ & $0.25$ & $1.050(7)$ & $-0.6219(2)$ & $0.014599$ & $0.018514$ & $0.428(7)$ & $1.076827$ & $-0.471531$ & $-0.177(7)$ \\
    \cline{1-10}
   $4.0$ & $0.25$ & $0.269(1)$ & $-0.3302(1)$ & $0.005571$ & $0.006558$ & $-0.061(1)$ & $0.269207$ & $-0.235766$ & $-0.094(1)$ \\
    \cline{1-10}
   $6.0$ & $0.25$ & $0.1240(8)$ & $-0.22838(6)$ & $0.003132$ & $0.003571$ & $-0.1044(9)$ & $0.119647$ & $-0.157177$ & $-0.0668(9)$ \\
    \cline{1-10}
   $8.0$ & $0.25$ & $0.0722(4)$ & $-0.17568(3)$ & $0.002072$ & $0.002319$ & $-0.1035(5)$ & $0.067302$ & $-0.117883$ & $-0.0529(5)$ \\
    \cline{1-10}
   $10.0$ & $0.25$ & $0.0490(2)$ & $-0.14324(2)$ & $0.001501$ & $0.001660$ & $-0.0942(3)$ & $0.043073$ & $-0.094306$ & $-0.0430(3)$ \\
    \cline{1-10}
   $40.0$ & $0.25$ & $0.00493(1)$ & $-0.039449(3)$ & $0.000198$ & $0.000207$ & $-0.03452(2)$ & $0.002692$ & $-0.023577$ & $-0.01363(2)$ \\
    \cline{1-10}
   $1.0$ & $0.5$ & $5.72(2)$ & $-1.088(1)$ & $0.036021$ & $0.052015$ & $4.63(3)$ & $5.973201$ & $-0.702678$ & $-0.64(3)$ \\
    \cline{1-10}
   $2.0$ & $0.5$ & $1.435(5)$ & $-0.5917(2)$ & $0.014563$ & $0.018516$ & $0.844(5)$ & $1.493300$ & $-0.351339$ & $-0.298(5)$ \\
    \cline{1-10}
   $4.0$ & $0.5$ & $0.367(1)$ & $-0.3206(1)$ & $0.005571$ & $0.006556$ & $0.046(1)$ & $0.373325$ & $-0.175669$ & $-0.151(1)$ \\
    \cline{1-10}
   $6.0$ & $0.5$ & $0.167(1)$ & $-0.2240(1)$ & $0.003131$ & $0.003570$ & $-0.057(1)$ & $0.165922$ & $-0.117113$ & $-0.106(1)$ \\
    \cline{1-10}
   $8.0$ & $0.5$ & $0.0954(5)$ & $-0.17331(6)$ & $0.002072$ & $0.002319$ & $-0.0779(6)$ & $0.093331$ & $-0.087835$ & $-0.0834(6)$ \\
    \cline{1-10}
   $10.0$ & $0.5$ & $0.0623(2)$ & $-0.14164(2)$ & $0.001501$ & $0.001660$ & $-0.0793(2)$ & $0.059732$ & $-0.070268$ & $-0.0688(2)$ \\
    \cline{1-10}
   $40.0$ & $0.5$ & $0.00556(1)$ & $-0.039373(3)$ & $0.000198$ & $0.000207$ & $-0.03382(2)$ & $0.003733$ & $-0.017567$ & $-0.01998(2)$ \\
    \cline{1-10}
   $1.0$ & $1.0$ & $9.72(2)$ & $-0.938(1)$ & $0.030389$ & $0.059999$ & $8.78(3)$ & $9.920268$ & $-0.438076$ & $-0.70(3)$ \\
    \cline{1-10}
   $2.0$ & $1.0$ & $2.419(5)$ & $-0.5280(4)$ & $0.013611$ & $0.019720$ & $1.891(6)$ & $2.480067$ & $-0.219038$ & $-0.370(6)$ \\
    \cline{1-10}
   $4.0$ & $1.0$ & $0.597(1)$ & $-0.2885(3)$ & $0.005475$ & $0.006666$ & $0.309(1)$ & $0.620017$ & $-0.109519$ & $-0.202(1)$ \\
    \cline{1-10}
   $6.0$ & $1.0$ & $0.272(1)$ & $-0.2107(3)$ & $0.003113$ & $0.003591$ & $0.061(1)$ & $0.275563$ & $-0.073013$ & $-0.142(1)$ \\
    \cline{1-10}
   $8.0$ & $1.0$ & $0.1524(5)$ & $-0.1647(1)$ & $0.002067$ & $0.002325$ & $-0.0123(7)$ & $0.155004$ & $-0.054760$ & $-0.1125(7)$ \\
    \cline{1-10}
   $10.0$ & $1.0$ & $0.0985(3)$ & $-0.1356(1)$ & $0.001500$ & $0.001661$ & $-0.0371(5)$ & $0.099203$ & $-0.043808$ & $-0.0925(5)$ \\
    \cline{1-10}
   $40.0$ & $1.0$ & $0.00728(2)$ & $-0.03891(1)$ & $0.000198$ & $0.000207$ & $-0.03163(3)$ & $0.006200$ & $-0.010952$ & $-0.02688(3)$ \\
    \cline{1-10}
   $1.0$ & $2.0$ & $18.17(2)$ & $-0.759(1)$ & $0.019490$ & $0.088388$ & $17.41(2)$ & $18.360596$ & $-0.241606$ & $-0.71(2)$ \\
    \cline{1-10}
   $2.0$ & $2.0$ & $4.550(5)$ & $-0.4444(3)$ & $0.009996$ & $0.026601$ & $4.106(5)$ & $4.590149$ & $-0.120803$ & $-0.363(5)$ \\
    \cline{1-10}
   $4.0$ & $2.0$ & $1.139(1)$ & $-0.2586(1)$ & $0.004619$ & $0.007888$ & $0.880(1)$ & $1.147537$ & $-0.060401$ & $-0.207(1)$ \\
    \cline{1-10}
   $6.0$ & $2.0$ & $0.5048(9)$ & $-0.1877(5)$ & $0.002806$ & $0.003981$ & $0.317(1)$ & $0.510017$ & $-0.040268$ & $-0.153(1)$ \\
    \cline{1-10}
   $8.0$ & $2.0$ & $0.2833(5)$ & $-0.1495(3)$ & $0.001932$ & $0.002487$ & $0.1338(8)$ & $0.286884$ & $-0.030201$ & $-0.1229(8)$ \\
    \cline{1-10}
   $10.0$ & $2.0$ & $0.1821(3)$ & $-0.1247(2)$ & $0.001432$ & $0.001739$ & $0.0574(5)$ & $0.183606$ & $-0.024161$ & $-0.1021(5)$ \\
    \cline{1-10}
   $40.0$ & $2.0$ & $0.01204(2)$ & $-0.03767(1)$ & $0.000197$ & $0.000208$ & $-0.02563(4)$ & $0.011475$ & $-0.006040$ & $-0.03106(4)$ \\
    \cline{1-10}
   $1.0$ & $4.0$ & $35.56(2)$ & $-0.5813(5)$ & $0.010552$ & $0.144170$ & $34.97(2)$ & $35.662094$ & $-0.125510$ & $-0.56(2)$ \\
    \cline{1-10}
   $2.0$ & $4.0$ & $8.882(4)$ & $-0.3558(2)$ & $0.005782$ & $0.045121$ & $8.526(5)$ & $8.915523$ & $-0.062755$ & $-0.326(5)$ \\
    \cline{1-10}
   $4.0$ & $4.0$ & $2.225(2)$ & $-0.2164(2)$ & $0.002962$ & $0.012259$ & $2.008(2)$ & $2.228881$ & $-0.031377$ & $-0.189(2)$ \\
    \cline{1-10}
   $6.0$ & $4.0$ & $0.9875(9)$ & $-0.1607(1)$ & $0.001943$ & $0.005743$ & $0.827(1)$ & $0.990614$ & $-0.020918$ & $-0.143(1)$ \\
    \cline{1-10}
   $8.0$ & $4.0$ & $0.5550(5)$ & $-0.1297(1)$ & $0.001419$ & $0.003384$ & $0.4254(6)$ & $0.557220$ & $-0.015689$ & $-0.1162(6)$ \\
    \cline{1-10}
   $10.0$ & $4.0$ & $0.3554(3)$ & $-0.1095(1)$ & $0.001102$ & $0.002261$ & $0.2459(4)$ & $0.356621$ & $-0.012551$ & $-0.0982(4)$ \\
    \cline{1-10}
   $40.0$ & $4.0$ & $0.02265(2)$ & $-0.03537(1)$ & $0.000188$ & $0.000217$ & $-0.01272(4)$ & $0.022289$ & $-0.003138$ & $-0.03187(4)$ \\
    \cline{1-10}
   $1.0$ & $8.0$ & $70.51(2)$ & $-0.4368(3)$ & $0.005390$ & $0.207543$ & $70.07(2)$ & $70.572060$ & $-0.063641$ & $-0.43(2)$ \\
    \cline{1-10}
   $2.0$ & $8.0$ & $17.625(4)$ & $-0.2655(1)$ & $0.003014$ & $0.083228$ & $17.360(4)$ & $17.643015$ & $-0.031821$ & $-0.251(4)$ \\
    \cline{1-10}
   $4.0$ & $8.0$ & $4.398(3)$ & $-0.1736(2)$ & $0.001604$ & $0.022492$ & $4.224(4)$ & $4.410754$ & $-0.015910$ & $-0.170(4)$ \\
    \cline{1-10}
   $6.0$ & $8.0$ & $1.955(1)$ & $-0.1320(1)$ & $0.001089$ & $0.010222$ & $1.823(1)$ & $1.960335$ & $-0.010607$ & $-0.127(1)$ \\
    \cline{1-10}
   $8.0$ & $8.0$ & $1.0986(9)$ & $-0.1080(1)$ & $0.000821$ & $0.005843$ & $0.991(1)$ & $1.102688$ & $-0.007955$ & $-0.104(1)$ \\
    \cline{1-10}
   $10.0$ & $8.0$ & $0.7039(6)$ & $-0.09220(9)$ & $0.000656$ & $0.003794$ & $0.6117(6)$ & $0.705721$ & $-0.006364$ & $-0.0877(6)$ \\
    \cline{1-10}
   $40.0$ & $8.0$ & $0.04429(3)$ & $-0.03196(2)$ & $0.000145$ & $0.000283$ & $0.01232(6)$ & $0.044108$ & $-0.001591$ & $-0.03020(6)$ \\
    \cline{1-10}
  \end{tabular}
  \caption{Measured energies at all densities and temperatures simulated for the polarized ($\xi=1.0$) gas. Shown for each density $r_{s}$ and temperature $\Theta$ are the size-corrected values for the Kinetic $K$ and Potential $V$ energies, their respective finite-size corrections $\Delta K_{N}$ and $\Delta V_{N}$, the resulting total energy $E_{tot}$, the free electron energy $E_{0}$, the Hartree-Fock exchange energy $E_{x,HF}$, and the resulting correlation energy $E_{c}$.}
  \label{tab:energies-1}
\end{table}

\begin{table}
  \begin{tabular}{|c|c|c|c|c|}
    \cline{1-5}
    $r_{s}$ & $E_{tot}(0)$ & $\displaystyle \lim_{T\rightarrow 0} E_{tot}(T)$ & $E_{c}(0) $ & $\displaystyle \lim_{T\rightarrow 0} E_{c}(T)$ \\
    \cline{1-5}
    $1.0$ & $2.2903(1)^{d}$ & $2.29(1)$ & $-0.0632(1)$ & $-0.07(1)$ \\
    \cline{1-5}
    $2.0$ & $0.2517(6)^{a}$ & $0.251(2)$ & $-0.0480(6)$ & $-0.048(2)$ \\
    \cline{1-5}
    $4.0$ & $-0.1040(1)^{d}$ & $-0.1042(6)$ & $-0.0346(1)$ & $-0.0348(6)$ \\
    \cline{1-5}
    $6.0$ & $-0.1230(1)^{d}$ & $-0.1228(3)$ & $-0.0280(1)$ & $-0.0278(3)$ \\
    \cline{1-5}
    $8.0$ & $-0.1134(1)^{d}$ & $-0.1130(2)$ & $-0.0239(1)$ & $-0.0235(2)$ \\
    \cline{1-5}
    $10.0$ & $-0.1013(1)^{a}$ & $-0.1013(1)$ & $-0.0209(1)$ & $-0.0209(1)$ \\
    \cline{1-5}
    $40.0$ & $0.0351348(7)^{c}$ & $-0.034894(8)$ & $0.0618048(7)$ & $-0.008224(8)$ \\
    \cline{1-5}
  \end{tabular}
  \caption{Zero-temperature extrapolations, $\displaystyle \lim_{T\rightarrow 0} E_{tot}(T)$, of finite-temperature PIMC calculations for the polarized ($\xi=1.0$). We compare $E_{tot}(0)$ directly to previous QMC studies where possible (a, \cite{PhysRevLett.45.566}), (b, \cite{PhysRevE.66.036703}), (c, \cite{PhysRevB.58.6800}), otherwise the Perdew-Zunger parameterization (d, \cite{PhysRevB.23.5048}) is used.}
  \label{tab:E0-1}
\end{table}

\begin{table}
  \begin{tabular}{|c|c|c|c|c|}
    \cline{1-5}
    $r_{s}$ & $\Theta$ & $\langle sgn \rangle$ &  $E_{tot}^{exact}$ & $E_{tot}$ \\
    \cline{1-5}
    $4.0$ & $0.0625$ & $-0.00055(62)$ & $-0.5(1)$ & $-0.1023(7)$ \\
    \cline{1-5}
    $10.0$ & $0.0625$ & $-0.002(1)$ & $-0.16(2)$ & $-0.1010(1)$ \\
    \cline{1-5}
    $1.0$ & $1.0$ & $0.0023(5)$ & $2(4)$ & $8.78(3)$ \\
    \cline{1-5}
    $4.0$ & $1.0$ & $0.0725(2)$ & $0.306(3)$ & $0.309(1)$ \\
    \cline{1-5}
    $10.0$ & $1.0$ & $0.4076(5)$ & $-0.0374(2)$ & $-0.0371(5)$ \\
    \cline{1-5}
    $40.0$ & $1.0$ & $0.9498(2)$ & $-0.031581(6)$ & $-0.03163(3)$ \\
    \cline{1-5}
    $1.0$ & $2.0$ & $0.0187(1)$ & $17.5(4)$ & $17.41(2)$ \\
    \cline{1-5}
    $1.0$ & $4.0$ & $0.1989(1)$ & $34.97(3)$ & $34.97(2)$ \\
    \cline{1-5}
    $1.0$ & $8.0$ & $0.5286(1)$ & $70.11(2)$ & $70.07(2)$ \\
    \cline{1-5}
    $4.0$ & $8.0$ & $0.8408(2)$ & $4.226(2)$ & $4.224(4)$ \\
    \cline{1-5}
    $10.0$ & $8.0$ & $0.93796(7)$ & $0.6115(1)$ & $0.6117(6)$ \\
    \cline{1-5}
    $40.0$ & $8.0$ & $0.99851(1)$ & $0.012275(9)$ & $0.01232(6)$ \\
    \cline{1-5}
  \end{tabular}
  \caption{Comparison of signful calculation $E_{tot}^{exact}$ with the fixed-node calculations $E_{tot}$ for the polarized ($\xi=1.0$) gas at select densities and temperatures. The average value of the sign is shown for reference.}
  \label{tab:Exct-1}
\end{table}




\end{document}